\documentclass[journal]{IEEEtran}
\ifCLASSINFOpdf
\else
\fi
\usepackage{url}

\usepackage{graphicx}
\usepackage{caption}

\begin{document}
%
\title{A Security Perspective on Unikernels}
%
%
%

\author{Joshua Talbot, Przemek Pikula, Craig Sweetmore, Samuel Rowe, Hanan Hindy,~\IEEEmembership{Member,~IEEE,} Christos Tachtatzis,~\IEEEmembership{Senior Member,~IEEE,} and Robert Atkinson,~\IEEEmembership{Senior Member,~IEEE,} Xavier Bellekens,~\IEEEmembership{Member,~IEEE}
\thanks{Manuscript received August 30, 2019.}}

%
%

\markboth{}%
{Talbot \MakeLowercase{\textit{et al.}}: A Security Perspective on Unikernels for IEEE Cloud Computing}

\maketitle

\begin{abstract}
Cloud-based infrastructures have grown in popularity over the last decade leveraging virtualisation, server, storage, compute power and network components to develop flexible applications. The requirements for instantaneous deployment and reduced costs have led the shift from virtual machine deployment to containerisation, increasing the overall flexibility of applications and increasing performances. However, containers require a fully fleshed operating system to execute, increasing the attack surface of an application. Unikernels, on the other hand, provide a lightweight memory footprint, ease of application packaging and reduced start-up times. Moreover, Unikernels reduce the attack surface due to the self-contained environment only enabling low-level features. In this work, we provide an exhaustive description of the unikernel ecosystem; we demonstrate unikernel vulnerabilities and further discuss the security implications of Unikernel-enabled environments through different use-cases.
\end{abstract}


\IEEEpeerreviewmaketitle

\section{Introduction}
Cloud computing is comprised of various virtualisation architectural models enabling users to build heterogeneous services comprised of multiple resources such as network devices, software components, serverless components, and containers. However, a new paradigm focusing on transient micro-services based on Unikernels has emerged and is becoming progressively popular. Unikernel's on-demand properties, low running costs and elasticity make it a perfect candidate for transient services. With the rise of multi-tenancy, multi-cloud infrastructure and the heterogeneity of the services proposed, the complexity of the ecosystem is constantly increasing, leading to a tremendous attack surface to cover and protect. In addition, the supply chain is often composted of various third party libraries, Operating Systems~(OS) and re-implemented operating system functions or legacy code enabled, to ensure retro-compatibility between systems. Existing work on Unikernels focuses mainly on its applications across a broad range of technologies, as well as, its integration with the host platform. However, while Unikernels claim a reduced attack surface, to the best knowledge of the authors, the security of Unikernels and their attack surface has not yet been explored in depth. In this manuscript, we review and explore Unikernel’ security ecosystems as the rise of transient microservices will make Unikernels prevalent in cloud infrastructure. The remainder of this manuscript is structured as follows; 

\section{Virtualisation, Containers and Unikernels}

\subsection{Virtualisation}
Virtualisation is the process of emulating a system, or multiple systems, using the resources of a host machine~\cite{Azure}. This can be used to re-create networks, or completely isolated machines, increasing the versatility of hardware. Virtualisation improves security through isolation as individual virtual machines cannot communicate with each other without explicitly specifying a connection. This isolation means that if the virtual machine administrator account is compromised, the attacker will not be able to access the host or other virtual machines running on it. This is facilitated by an additional level of user-privilege on the host that controls the guest~\cite{VMware1}. 

\subsubsection{Virtualisation Types}
There are a variety of virtualisation types, each with their own advantages and disadvantages. 

\paragraph{Full Virtualisation}
This type of virtualisation virtualises the hardware the guest machines runs on. This can either be hardware-assisted, with the hardware itself supporting the virtualisation or software-assisted, where  the operating system interfaces with the hardware~\cite{SuperUser}. The former type of virtualisation's main appeal is its ability to emulate hardware, allowing for consistent performance, improved reliability, and isolation. If an attacker gained control over the machine he would have no knowledge of the real hardware of running on the host. While the malicious user might not be able to interact with the host environment, he might, however, be able to discover that he is interacting with a virtualised environment~\cite{virtual_attack}. This, in turn, will allow the attacker to fine-tune his attacks to target the virtual machine itself, reducing its security.

\paragraph{Para-Virtualisation}
Though largely antiquated (having support removed from the Linux kernel in 2009~\cite{linKernPara}), this type of virtualisation ensures that the virtual machine interacts with a software interface instead of the hardware directly. This allows the virtual machine to use Application Programming Interfaces~(APIs) to make system calls that would otherwise be hard to virtualise. This improves the overall efficiency since, the most complex system calls are abstracted through the API~\cite{VMTech}. However, with the improving efficiency of hardware virtualisation, para-virtualisation no longer provides tangible performance benefits~\cite{linKernPara}.

\paragraph{OS-Virtualisation}
OS-Virtualisation is where a single kernel can run multiple occurrences of the operating system as containers, each of which acting as an isolated machine. These containers place less emphasis on recreating an entire machine, but rather focus on the user space, allowing users to run multiple operating systems and associated software on a single machine for convenience.  The containers do not have access to the hardware of the physical machine, and will typically use the same OS as the host, which in turn can limit the application of the machine~\cite{UnixArena}. This also implies that if the kernel is ever compromised, all associated containers will be compromised~\cite{jaxenter}.

\begin{figure}[thb]
    \centering
    \captionsetup{justification=centering,margin=2cm}
    \includegraphics[scale=.65]{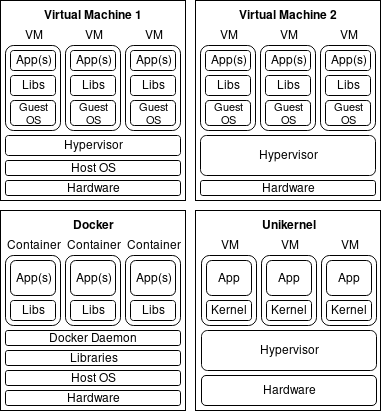}
    \caption{Virtualisation Types}
    \label{fig:architecture}
\end{figure}

\begin{figure*}[htb]
    \centering
    \includegraphics[width=0.7\textwidth]{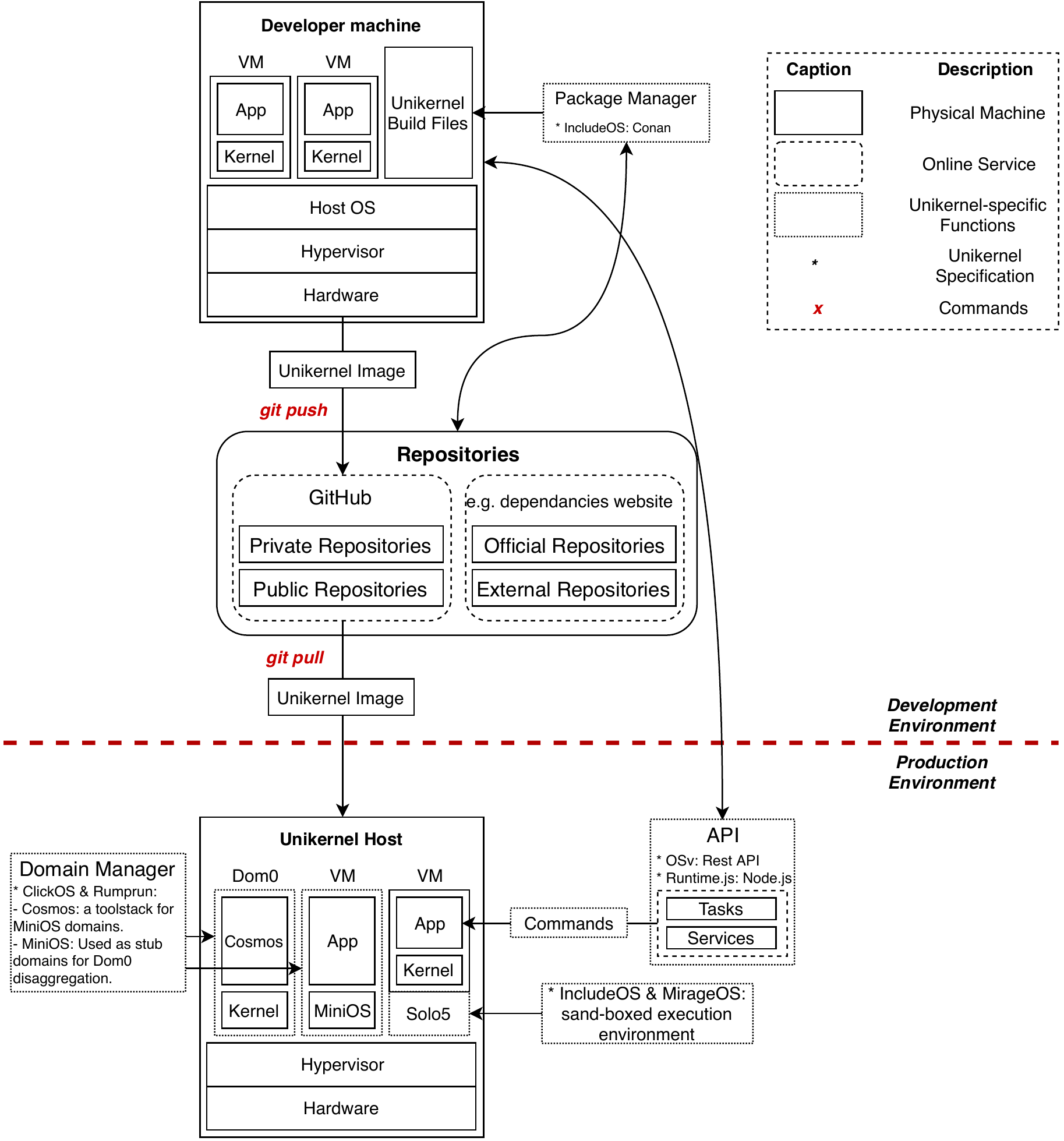}
    \caption{Unikernel Ecosystem}
    \label{fig:ecosystem}
\end{figure*}

\subsection{Docker}
Docker is software enabling OS-Virtualisation through the use of the `docker engine' which manages the containers on the host. Docker, like most OS-Virtualisation, opts for process-level isolation over full isolation. Whilst this makes it more efficient for running isolated applications, Docker systems demonstrate numerous vulnerabilities, as highlighted in~\cite{VulnDocker}.

\subsection{Unikernel}
\subsubsection{Unikernel Types}
There are two major kinds of Unikernels, whose security profiles differ slightly: Clean Slate and Legacy~\cite{Devconf.cz}. 



\paragraph{Clean Slate}
Clean slate Unikernels do not try to emulate classical OS in any regards. They are written in a single programming language and provide interfaces for external communications (i.e. networking) in the same language. Examples include MirageOS (OCaml), IncludeOS (C++), HalVM (Haskell), LING (Erlang) and runtime.js (JavaScript).
Clean slate Unikernels allow language specific virtual machines, like the Java Virtual Machine~(JVM), to function as actual virtual machines. C libraries, while present, are transparent to userspace code. The most straightforward implementation can be found in runtime.js, which wraps Chromium's V8 Javascript engine inside a lightweight kernel~\cite{JsOS}.

\paragraph{Legacy}
Legacy Unikernels, on the other hand, implement a subset of POSIX to ensure unmodified software to run, while some only require minor configuration changes. They don't support timesharing (the ability to simultaneously run multiple independent programs), instead, they delegate this role to the virtualisation layer.
Unikernels such as OSv and Graphene focus on ensuring Linux compatibility and software interoperability, re-implementing system call interfaces, while the Rumprun unikernel implements a subset of FreeBSD's syscalls~\cite{Graphene}~\cite{tsai2014cooperation}.

Figure~\ref{fig:architecture} summarises the primary differences between the different types of virtualisation including unikernels. Figure~\ref{fig:architecture} (top-left) shows the layout of software assisted virtualisation, and OS-virtualisation. The virtualisation is run on top of the OS by a hypervisor, which can allow for the creation of virtual hardware. Figure~\ref{fig:architecture} (top-right) shows hardware-assisted virtualisation, and para-virtualisation. The hardware itself (sometimes assisted by the hypervisor) runs the virtual machines itself, while Figure~\ref{fig:architecture} (bottom-left) demonstrates how containers are implemented and finally Figure~\ref{fig:architecture} (bottom-right) provides a overview of Unikernels virtualisation architecture. As demonstrated, Unikernels do not require an operating system to function correctly.

\subsubsection{Isolation}
Software running on a Unikernel is less isolated from the hypervisor than software running on a virtual machine, but more isolated than software running in a container. While Unikernels rely on the hypervisor for their isolation, they also bring their own kernel with a reduced attack surface. This bespoke kernel makes Unikernels more isolated than containers. However, similar to any software, it is up to the Unikernel developers to supply Unikernels with intrinsic security.

\subsection{Ecosystem}
Figure~\ref{fig:ecosystem} shows how the development and production environments differ, along with the Unikernel's distribution. In development, an OS would be required in setting up the Unikernel and its features, while in production it is optimal to deploy the Unikernel without a host OS for increased optimisation and security.
This can be managed through an online repository to host and modify the Unikernel image. Moreover, the environment/functionalities vary depending on the Unikernel's distribution, which has been generalised in the example above. Features include a package manager (e.g. Conan) for building and downloading Unikernel compatible applications, a toolstack/domain manager (e.g. Cosmos) to manage unprivileged domains, and an API (e.g. Rest) to manage Unikernels remotely through issuing commands.

\subsection{Unikernel Usage}
Unikernels are primarily used in cloud computing, however, they also show potential in Internet of Things~(IoT) and networking devices. These platforms benefit from Unikernels flexible, lightweight and scalable framework. For cloud computing and networking devices, Unikernels prove more effective at utilising the available hardware, allowing for increased scalability in a more isolated, heterogeneous computing environment with a more optimised code base. Due to their efficiency, Unikernels also come with the benefit of much faster boot times, allowing for downed services or network nodes to be quickly restored with minimal overhead. The most compelling reason to use Unikernels is their potential for security which takes advantage of their reduced attack surface, isolation and, depending on the distribution, a robust set of security features~\cite{nanovms.com}. IoT devices also largely benefit from Unikernels lightweight code-base and scalability, allowing for a more complex and diverse set of services to be deployed on-demand with very low overhead, despite the hardware limitations.

    
\begin{figure*}
    \centering
    \includegraphics[width=0.5\textwidth]{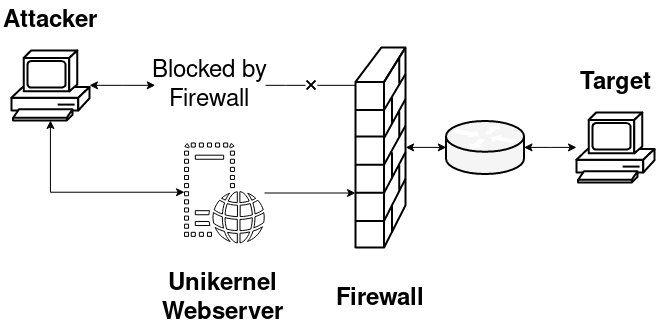}
    \caption{Pivoting Attack}
    \label{fig:pivot-mono}
\end{figure*}

\section{Unikernels Security Overview}
By limiting the code base of deployed applications, Unikernels inherently have a small and unique attack surface making them relatively secure. This is further achieved in some implementations by evaluating and modifying the implemented code~\cite{includeos.org-secure}, where developers are able to focus on hardening security to effectively mitigate existing attack vectors.

\subsubsection{Shell}
Numerous Unikernels do not implement a shell natively, making most types of payloads, that typically rely on bash, ineffective. Further preventing automated attacks or less experienced attackers from successfully exploiting a vulnerability by increasing the complexity of the payloads.

\subsubsection{System Calls}
System calls are often removed, or are not supported by Unikernels, hence,  malicious users are required to know the exact memory layout in order to invoke a function call such as \texttt{open()} or \texttt{write()}. The attack surface is further reduced through implementing randomised memory layouts at every build~\cite{includeos.org-secure}.
    
\subsubsection{Hardware Emulation}
By not emulating hardware interfaces such as floppy drives, Peripheral Component Interconnect~(PCI) bus or Graphics Processing Unit~(GPU), possible breakouts such as the 2015 Venom attack that affected QEMU can be prevented~\cite{CVE-QEMU}.  Unikernels such as IncludeOS and Mirage-OS are also able to mitigate this through running on solo5; a sandboxing interface between the Unikernels and the hypervisor. Solo5 allows for a minimal code base, removing code that would otherwise be unnecessary to cloud computing. This is achieved during compilation by determining the required dependencies from the imported libraries, application code and configuration files, further reducing the Virtual Machine Monitor~(VMM) overall attack surface.

\vspace{3mm}
Many Unikernel implementations rely solely on their reduced attack surface from the previously mentioned features or lack thereof, not taking into consideration known vulnerabilities relating to their existing attack vectors, such as minimal security features in Address Space Layout Randomisation~(ASLR)~\cite{nccgroup.trust}. This has led to criticism and debate discussing their design; The lack of separation between the user and kernel-space comes with great security concerns. For example, a successful buffer overflow attack on the Unikernels limited functionality  could give an attacker a foothold into kernel space, making privilege escalation unnecessary, as well as, code execution and pivoting to another target a potential threat. Furthermore, Figure~\ref{fig:pivot-mono}) shows an attacker blocked by the target's firewall can leverage a web-application running on a unikernel to exploit the network. By exploiting the web-application, the Unikernel API or through compromising the Unikernel itself, the attack could used  kernel functions to forge malicious packets, for denial of service, or to target other devices on the network enabling him to pivot~\cite{nccgroup.trust}. Hence, while demonstrating potential for scalability and improved security by reducing their attack surface, some Unikernels are yet to implement key security features.

\subsection{Immutable Infrastructure}
In traditional infrastructure, patches can be applied to upgrade packages, change the servers configuration and modify or upload code. This, however, poses the threat of malicious modifications being made by an attacker. Having an immutable infrastructure mitigates this by employing the ``destroy and provision" approach of needing to rebuild the Unikernel to make any changes. This not only prevents malicious modifications but also reduces the overall code complexity of having outdated configurations that could lead to new bugs and vulnerabilities.

\paragraph{Para-Virtualisation}
In some Unikernel implementations, para-virtualisation is implemented to restrict privileged operations to the hypervisor through an API. This essentially allows applications to run in ring 3 rather than ring 0, isolating the Unikernels application from the hardware level. Therefore, the hypervisor is able to enforce Write xor Execute (W\char`\^X) to its page tables by making executable pages immutable.

\paragraph{Heterogeneous Networking}
Increased heterogeneity can be achieved through either enabling multiple instances of a certain application to run with varying configurations and libraries or by using a separate Unikernel instance for each function in a network. For example, this can be implemented through using a set of Unikernels for hosting databases, each with their own protections and privileges depending on the sensitivity and nature of their data, rather than having a singular database that stores all of this data. Similarly, by allocating specific Unikernels to run the server, web service, etc\ldots, the network's architecture can be isolated to its critical functions, preventing possible knock-on effects of certain attacks, such as DDoS, from compromising the entire network and allowing for the affected elements of the network to be isolated and easily identifiable.

\subsection{Entropy}
Unikernels often have a low entropy as the hardware is virtualised, hence, randomly generated values persist across reboots, meaning, that if an attacker were to crash a Unikernel, they may be able to determine ramdomized values even after rebooting. This lack of entropy can lead to security features such as ASLR, stack cookies, TCP sequence numbers and access tokens, etc\ldots becoming ineffective~\cite{nccgroup.trust}. In Rumprun and MirageOS, this has been effectively mitigated through the implementation of RDRAND, however, since this is a common issue among vitalised platforms, this vulnerability requires to be checked on a platform basis before use~\cite{mirage.io}. Enforcing entropy persistence amongst Unikernels may affect incohesive generated Unikernels by having duplicate values. To mitigate this, generated seeds should be validated as non-duplicates otherwise, they may be vulnerable to nonce reuse attacks.

\section{Unikernel Weak Links}

\begin{figure*}[thb]
    \centering
    \captionsetup{justification=centering,margin=2cm}
    \includegraphics[scale=1.2]{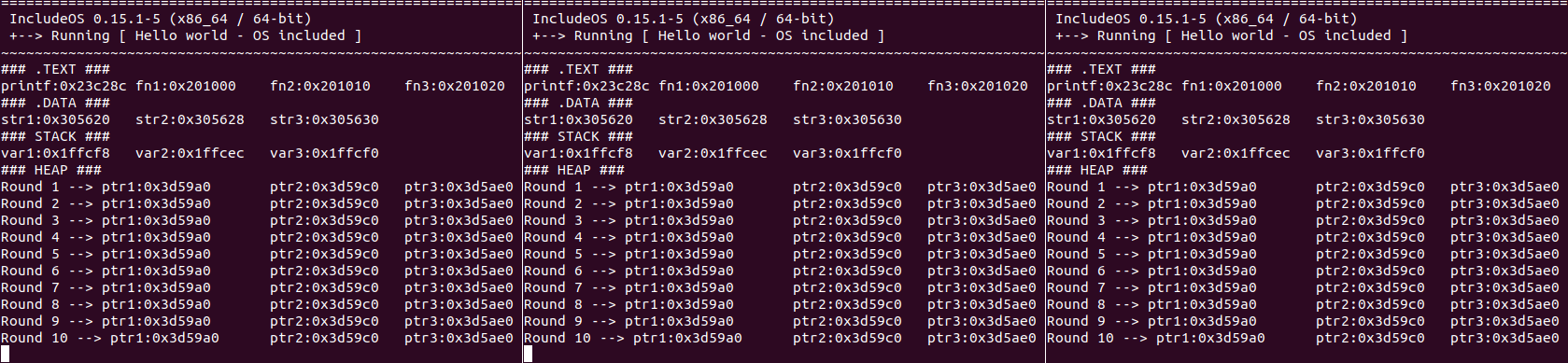}
    \caption{ASLR Test Results}
    \label{fig:ASLR}
\end{figure*}

\subsection{ASLR Vulnerability}
IncludeOS mentions its use of ASLR, however, it has been demonstrated that their implemention is either flawed or a bogus claim~\cite{nccgroup.trust}. In this section we replicate a successful attack against ASLR and the lattest IncludeOS version.   The ASLR attack is performed against latest stable release of Ubuntu (18.04.2 LTS) and IncludeOS (v0.15.0). The cloud Hello World demo from IncludeOS's GitHub repository was used as a baseline to host the ASLR testing service. This service prints the memory addresses of strings, the output of functions and variables on the stack, as well as, pointers to newly added blocks of memory onto the heap. To verify the assertion of ASLR being implemented after each build the tests were repeated after deleting and then rebuilding the Unikernels files.

\subsection{Results}
Figure~\ref{fig:ASLR} illustrates the results showing that ASLR was not implemented in the latest version of IncludeOS, which is evidenced by the unchanging memory addresses of the stored values.  Furthermore, Rumprun, IncludeOS and MirageOS versions of Unikernels have been tested and confirmed to not implement ASLR, page protections or stack canaries, and set their memory to RWX and hence are vulnerable against multiple attacks. 

\subsection{Unikernel Limitations}
\paragraph{Protection Rings}
In traditional operating systems, protection rings are used to set increasing levels of access to the operating system. The issue with Unikernels is that they run their applications in Ring 0 or the ``kernel ring". Specifically, one of the core ideas behind the structure of a Unikernel is that the kernel, operating system and application is contained in a single system~\cite{newstack.io}. One of the factors of this is that the Unikernel does not have the capability to provide additional protection rings, and so must go without. 
\paragraph{Guard Pages} unmapped pages between memory allocations used to cause segmentation faults are not implemented on these Unikernels either~\cite{nccgroup.trust}.

\paragraph{Debugging Tools}
Unikernel deployment is limited due to the increased difficulty in debugging . This is due to the removal of components of the operating system; standard commands for debugging such as netstat, tcpdump and ping are not present on the Unikernel~\cite{joyent.com, morethanseven.net}. Making it difficult for developer  maintaining and updating Unikernels as there is no easy maner to determine an issue in the code and would likely be required to perform trial-and-error testing consuming time and resources. An additional issue is that Unikernels cannot be updated while running, and require to be shut down, updated, rebuilt and run again for changes to take effect.

The combination of these factors creates the potential for the exploitation of a buffer overflow vulnerability to directly overwrite the instructions of the program. In turn, this allows for remote code execution, effectively granting the attacker remote access to the system, which will be met with administrator privileges due to the lack of protection rings.\newline

\subsection{Mitigations}
\begin{itemize}
    \item Entropy: Enforcing entropy persistence amongst Unikernels may affect groups of frequently generated Unikernels by sharing duplicate values. To mitigate this generated seeds should be validated as non duplicates otherwise they may be vulnerable to nonce reuse attacks.
    
    \item Hardening: A method of mitigating the chance of the Unikernel being compromised is to implement hardening.
    Hardening is where efforts are made to restrict what vulnerable parts of the system can be accessed by an attacker. Common techniques for this are consistent patching of the operating system and application, closing of unused ports, enforcement of password complexity and removal of default accounts~\cite{owasp.org}. 
    
    \subsubsection{Host Hardening}
    Host hardening makes it more difficult to exploit the applications running on the host. This can be accomplished through:
    \begin{itemize}
    \item reducing the attack surface
    \item protecting against stack overflows
    \item randomising memory layout (ASLR)
    \item protecting against buffer overflows
    \item encrypting data wherever possible
    \end{itemize}
    
    Unikernels, at least in theory, provide the ultimate attack surface reduction. How much the surface is reduced varies by Unikernel. However, in general, clean slate Unikernels reduce the attack surface more then legacy Unikernels do. This is because compiling the application and the kernel it will run on into a single program allows the compiler to verify all methods data is passed around. Interfacing directly within a typesafe, high level programming language allows clean slate Unikernels a vastly reduced attack surface compared to passing data through pipes and C library functions. This reduces the risk of buffer overflows.

    \subsubsection{Library Hardening}
    For Unikernels, there are two very different forms of library hardening:
    - C standard library hardening, which affects legacy Unikernels and, to a lesser degree, clean slate Unikernels,
    - Native library hardening, which affects clean slate Unikernels. However, unlike traditional OS where C is the native language, a clean slate Unikernel the language the Unikernel was implemented in is the native.

    Unikernels use many different C standard libraries. OSv uses musl, IncludeOS uses newlib, rumprun uses libc from NetBSD and UKL uses glibc~\cite{OSv}~\cite{nccgroup.trust}~\cite{UKL}. NCC testing has revealed that some, like newlib, lack support for the \textunderscore  FORTIFY\textunderscore SOURCE macro that can be used to automatically detect some buffer overflows in common C functions. The lack of  \textunderscore FORTIFY\textunderscore SOURCE forces developers to manually verify bounds checks in all the programs. Other parts of the Standard C library, that are often exploitable and should be hardened include format specifiers.

    Native library hardening depends on both how robust the Unikernel's native programming language standard library is, and how securely are the Unikernel specific wrappers over external interfaces implemented.

    \subsubsection{Networking}
    Network hardening usually involves disabling unnecessary services. Unikernels come with few, if any, making them relatively tricky to crack out of the box. However, the network services they come with can be quite vulnerable if left exposed. For example, in OSv the REST API used to control the Unikernel can replace the command line, read and write files and directories. Exposing it to attackers gives them command execution, Local and Remote File Inclusion.
    
    \subsubsection{Security Modules and their comparison with Traditional Linux Host Hardening} 
    There are several potential security modules that could be installed and implemented by an administrator. All provided examples making use of a system for Mandatory Access Controls, where the OS will control the ability for individual users to grant or deny access to resource objects on a file system~\cite{searchsecurity.techtarget.com}
    These are: SELinux (Security-Enhanced Linux), which provides a system for Mandatory Access Controls and defines the access and transition rights of all users, applications, processes and files on the system. \cite{access.redhat.com}, AppArmor, which binds the access control attributes to the programs instead of the users, which will either report or outright prevent access from chosen profiles \cite{wiki.ubuntu.com} 
\end{itemize}

\section{Conclusion}
Cloud environments are constantly evolving to reduce deployment costs and decrease the complexity of virtualised solutions. Throughout this paper we presented an overview of the strength and weaknesses of Unikernels. We demonstrated that some Unikernels are vulnerable to known attacks such as buffers overflows and did not yet integrated best practices to alleviate common vulnerabilities. 

\bibliographystyle{IEEEtran}
\bibliography{references}



\end{document}